# OpenAlex: A fully-open index of scholarly works, authors, venues, institutions, and concepts

Jason Priem*, Heather Piwowar*, Richard Orr*

*jason@ourresearch.org; heather@ourresearch.org; richard@ourresearch.org
OurResearch, 500 Westover Dr #8234, Sanford, NC, 27330 (USA)

## Introduction

In May 2021, Microsoft announced that it was discontinuing support for Microsoft Academic Graph (Sinha, Shen, et al., 2015), a free and widely-used Scientific Knowledge Graph (SKG). This was met with considerable concern, as MAG was viewed as difficult to replace with existing systems (Tay, Martín-Martín, and Hug, 2021). The OpenAlex project was created to address this concern. It launched as a drop-in replacement for MAG contemporaneously with MAG's retirement on January 1st 2022.

Although still in its nascency, as a fully-open (100% open data, open API, open-source code) source of scholarly metadata, OpenAlex has potential to improve the transparency of research evaluation, navigation, representation, and discovery, adding to the growing list of other open and partly-open SKGs such as OpenCitations (Peroni, Shotton, & Vitali, 2017), AMiner (Tang, Zhang et al., 2008), PID Graph (Fenner & Aryani, 2019), Open Research Knowledge Graph (Jaradeh, Oelen et al., 2019), Semantic Scholar (Ammar, Groeneveld et al., 2018), and the OpenAIRE research graph (Manghi, Atzori et al., 2019).

## Schema and IDs

The OpenAlex dataset is a heterogeneous directed graph, composed of five types of scholarly entities, and the connections between them.

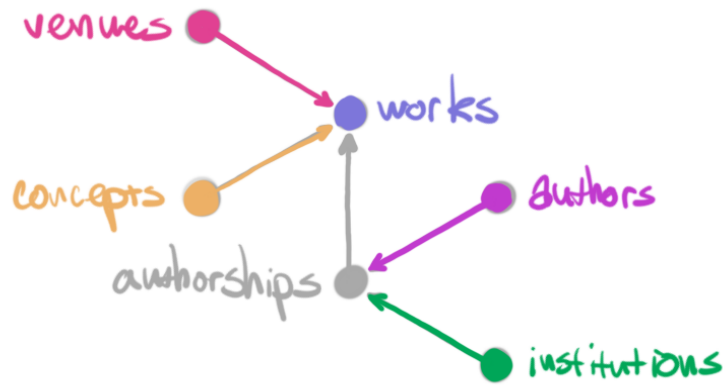

*Figure 1: Sketch of the OpenAlex graph data model.*

All entities are assigned a persistent OpenAlex ID, which acts as a primary key in the dataset. This ID is expressed as a URL, which can resolve to either a human-readable (webpage) or machine-readable (JSON object) representation. Where possible, entities are also assigned IDs from external systems, in order to increase interoperability. Particular effort is given to assinging a Canonical External ID (CEID). The CEID varies by entity type, and was selected based on the extent of its adoption by the community.

**The OpenAlex entities**
There are five types of entities in OpenAlex: works, authors, venues, institutions, and concepts:

*Works*
Works are scholarly documents like journal articles, books, datasets, and theses. Although all entities are "first-class citizens" in the OpenAlex graph, works are particularly important, because connection to scholarly works is what makes authors, venues, institutions, and concepts "scholarly." Much of the information about the links between works and other entities is obtained by parsing work metadata–either in structured form (eg, in the Crossref API) or unstructured form (eg, on a publisher landing page).

 OpenAlex indexes about 209M works, with about 50,000 added daily. The CEID for works is DOI; about half of works have one. We collect new works from many sources, including Crossref, PubMed, institutional and disciplinary repositories (eg, arXiv). In addition to these sources, MAG is a key source data source for older works.

*Authors*
Authors are defined as people who create works. OpenAlex indexes about 213M authors, with thousands added daily. The CEID for authors is ORCID; only a small percentage of authors have an ORCID, but the percentage is higher for authors of more recent works. OpenAlex uses ORCID as a feature to help algorithmically disambiguate authors, when it is available. The system also uses authors' publication records and citation histories in the disambiguation algorithm. Authors are connected to works via the "authorship" object, which formalises the three-way claim customarily expressed in work metadata: "author $A$, affiliated with institution(s) $I$, is a creator of work $W$."

*Venues*
Venues are defined as places that host works. OpenAlex indexes about 124,000 venues. There are several types, including journals, conferences, preprint repositories, and institutional repositories. The CEID for venues is ISSN-L or "linking ISSN." The ISSN-L is a single ISSN that groups all the ISSNs a particular publication may have . About 90% of venues in OpenAlex have an ISSN-L or ISSN. Works are often hosted in multiple venues, often in multiple versions. For example, a work may live in preprint form on arXiv, but as a version of record (VoR) on a publisher's webpage. OpenAlex uses a fingerprinting algorithm to match these two versions, and report them both, flagging the VoR as the primary host. The system also automatically determines the version and licence of all copies, where possible.

*Institutions*
Institutions are organisations to which authors claim affiliations. OpenAlex indexes about 109,000 institutions. The CEID for institutions is the ROR ID; about 94% of institutions in OpenAlex have ROR IDs. In order to link institutions to works, we parse every affiliation listed by every author; these strings are obtained from both structured sources (eg, PubMed) and unstructured ones (publisher webpages). We extract and normalise institution strings from affiliation statements, then link them to ROR IDs where possible; this is done using a two-step algorithm that combines both rules-based and machine-learning stages. Like authors, institutions are linked to works via an "authorship" object.

*Concepts*
Concepts are abstract ideas that works are about. OpenAlex indexes about 65k concepts. The CEID for OpenAlex concepts is the Wikidata ID, and all concepts have one, because all OpenAlex concepts are also Wikidata concepts. Concepts are hierarchical; there are 19 root-level concepts, and 5 layers of descendents branch out from them. The concept tree, which is a modified version of the one used in MAG, contains about 65 thousand concepts in total. Works are assigned concepts based on their titles and abstracts, using an automated classifier that was trained on MAG's corpus. The code and models behind this classifier are open-source. Around 85% of works have at least one concept assigned.

**Open distribution**
There are three ways to obtain data from OpenAlex: a full data dump (updated fortnightly), a REST API (updated daily), and a web-based GUI (built on our own REST API). All these are free and require no registration or permission. The REST API has no rate-limits, although for high query loads (more than 100k/day) the data dump is recommended instead. The code behind OpenAlex is fully Open Source, and is available on the OurResearch GitHub account (https://github.com/ourresearch). OpenAlex is a project of OurResarch, a nonprofit devoted to open scholarly principles and an early adopter of The Principles of Open Scholarly Infrastructure (POSI; Bilder, Lin, & Neylon, 2020) which provide guidance for sustainably open development.

**Limitations and future work**
The OpenAlex project is still quite young, and there are many areas for improvement. Foremost is continued improvement in the parsing, normalisation, and disambiguation of entities, especially authors and institutions. This is particularly important given the real-world implications of this metadata, which may (rightly or wrongly) be used in high-stakes evaluation contexts. The dataset currently lacks metadata about funding sources, as well as metadata about corresponding authors. Overall, much work remains to be done in validating and studying completeness and accuracy of the dataset, particularly in comparison to similar tools. However, by leveraging, extending, and opening MAG's work, OpenAlex offers a promising alternative to toll-access data sources in this space.

**Acknowledgements**
Development of OpenAlex is funded by a grant from Arcadia: A charitable fund of Lisbet Rausing and Peter Baldwin. Thomas Scheidsteger and Robin Haunschild provided very helpful feedback on an early draft.

**References**
Ammar, W., Groeneveld, D., Bhagavatula, C., Beltagy, I., Crawford, M., ... Etzioni, O. (2018). Construction of the literature graph in semantic scholar. Proceedings of the 2018 Conference of the North American Chapter of the Association for Computational Linguistics: Human Language Technologies, Volume 3 (pp. 84–91). Association for Computational Linguistics. https://doi.org/10 .18653/v1/ N18-3011

Bilder G, Lin J, Neylon C (2020), The Principles of Open Scholarly Infrastructure, retrieved 29 April 2022, https://doi.org/10.24343/C34W2H

Fenner, M., & Aryani, A. (2019). Introducing the PID graph. https:// doi.org/10.5438/ JWVF-8A66

Jaradeh, M. Y., Oelen, A., Farfar, K. E., Prinz, M., D'Souza, J., ... Auer, S. (2019). Open research knowledge graph: Next genera- tion infrastructure for semantic scholarly knowledge. In Proceed- ings of the 10th International Conference on Knowledge Capture (pp. 243–246). New York: Association for Computing Machinery. https://doi.org/10.1145/3360901.3364435

Manghi, P., Atzori, C., Bardi, A., Schirrwagen, J., Dimitropoulos, H., ... Summan, F. (2019). OpenAIRE Research Graph Dump. Zenodo. https://doi.org/10.5281/zenodo.3516918

Peroni, S., Shotton, D., & Vitali, F. (2017). One year of the Open- Citations Corpus. In C. d'Amato et al. (Eds.), The semantic web – ISWC 2017 (pp. 184–192). Cham: Springer. https://doi.org/10 .1007/978-3-319-68204-4_19

Sinha, A., Shen, Z., Song, Y., Ma, H., Eide, D., ... Wang, K. (2015). An overview of Microsoft Academic Service (MAS) and applica- tions. In Proceedings of the 24th International Conference on World Wide Web (pp. 243–246). New York: Association for Computing Machinery. https://doi.org/10.1145/2740908 .2742839

Tang, J., Zhang, J., Yao, L., Li, J., Zhang, L., & Su, Z. (2008). Arnet- Miner: Extraction and mining of academic social networks. In Proceedings of the 14th ACM SIGKDD International Conference on Knowledge Discovery and Data Mining (pp. 990–998). New York: Association for Computing Machinery. https://doi.org/10 .1145/1401890.1402008

Tay, Aaron, Martín-Martín, Alberto and Hug, Sven E. (2021) Goodbye, Microsoft Academic – hello, open research infrastructure? Impact of Social Sciences Blog (27 May 2021). Retrieved 30 April, 2022, http://eprints.lse.ac.uk/id/eprint/111325